\documentstyle[12pt]{article}
\input epsf
\begin{document}
\newcommand{\etap}{\eta^\prime}
\newcommand{\vsig}{\mbox {\boldmath $\sigma$\unboldmath}}
\newcommand{\vep}{\mbox {\boldmath $\epsilon$\unboldmath}}
\bibliographystyle{unsrt}
\title{ Meson Photoproductions Off Nucleons In The 
Chiral Quark Model}
\author{ Zhenping Li \\
Physics Department, Peking University \\
 Beijing 100871, P.R.China\thanks{ 
 E-mail: ZPLI@ibm320h.phy.pku.edu.cn}}

\maketitle

\begin{abstract}
The meson photoproductions off nucleons in the chiral quark model are
described.  The role of the S-wave resonances in the second resonance region is
discussed, and it is particularly important for the Kaon, $\eta$ and $\eta'$ 
photoproductions.
\end{abstract}

Meson photoproductions of nucleons have always been a very important 
field to study the structure of hadrons.  It was the early
 investigations by Copley, Karl and Obryk\cite{cko} and Feynman, 
Kisslinger and Ravndal\cite{fkr} in the pion photoproduction that
provided the first evidence of an underlying $SU(6)\otimes O(3)$
symmetry for the baryon structure in the quark model.
 Extensive calculations 
and discussions of the photoproduction of baryon resonances later 
have not changed the conclusion in Refs. 1 and 2 
significantly.  These calculations in the framework of the quark
models have been limited on the transition amplitudes that are 
extracted from the photoproduction data by the phenomenological 
models, thus it is less model independent.
The challenge is whether one could go one step 
further to confront the photoproduction data directly with the 
quark model.  Such a step is by no means trivial, since it requires
that the transition amplitudes in the quark model  have correct 
off-shell behavior, which are usually evaluated on shell.  More 
importantly,  it also requires that the model with explicit quark 
and gluon degrees of freedom give a good description of the 
contributions to the photoproductions from the non-resonant 
background,  which are usually used to evaluate the contributions 
from s-channel resonances.  
The low energy theorem in the threshold pion photoproduction\cite{cgln}
is a crucial test in this regard, which the non-resonant contributions
dominate in the threshold region.  Our 
investigation\cite{zpli94} showed that the simple quark model is
no longer sufficient to recover the low energy theorem, and one has 
to rely on low energy QCD Lagrangian so that the meson baryon 
interaction is invariant under the chiral transformation.  During 
the past three years,  we have extended it to the kaon\cite{zpli951} 
and $\eta$\cite{zpli952} photoproductions by combining the 
low energy QCD Lagrangian and the quark model, and the initial 
results showed very good agreements between the theory and 
experimental data  with far less parameters.  In the remaining 
part of this paper,  I would like to review the quark model 
framework for the meson photoproductions, and discuss the important
role of s-wave resonances, in particular, those in the second 
resonance region in the $K$, $\eta$ and $\eta'$ photoproductions.

We shall discuss briefly the chiral quark model approach to the meson
photoproductions of the nucleon, as the detailed formalism in the quark 
model has been given in Refs. 5 and 6.  There are 
four components for the photoproductions of the pseudo-scalar mesons based 
on the low energy QCD Lagrangian in Ref. 7; the contact term and
the s-, u- and t- channel contributions, thus the matrix element
for the meson photoproductions can be written as 
\begin{equation}\label{11}
{\cal M}_{if}={\cal M}_{c}+{\cal M}_t+{\cal M}_s+{\cal M}_{u}.
\end{equation}
The contact term ${\cal M}_c$ in Eq. \ref{1} is generated by the gauge 
transformations of the axial vector in the QCD Lagrangian.  It is
proportional to the charge of the outgoing mesons, therefore, it does
not contribute to the productions of the charge neutral mesons, such as
the $K^0$ productions in the reactions $\gamma N\to K \Sigma$. Moreover,
the integrations of the spatial wavefunctions of the initial and final
baryons generate a form factor that has a maximum value at the forward 
angle and decreases as the scattering angle between the incoming 
photon and the outgoing meson increases.  This leads to an interesting
prediction from the quark model; the charged meson productions should be 
forward peaked above the threshold because of the dominance of the 
contact term in the low energy region.  The data in charged kaon 
and the neutral $\eta$ productions are quite consistent with this 
conclusion.  The second term ${\cal M}_t$ in Eq. \ref{11}
 is the t-channel $K^+$ 
exchange, and it is proportional to the charge of the outgoing mesons
as well.  This term is required so that the total transition 
amplitude in Eq. \ref{1} is gauge invariant.  The other t-channel 
exchanges, such as the $K^*$ and $K1$ exchanges in the kaon productions,
which played an important role in Ref. 8, are excluded with the
input of the duality hypothesis\cite{dolen,wjs}.  This was not imposed 
in our early investigation\cite{zpli951} of the kaon photoproductions,
in which the $K^*$ exchange was included.  

The u-channel contributions ${\cal M}_u$ in Eq. \ref{11} include $\Sigma$
($\Lambda+\Sigma^0$) exchanges for the $\Sigma^{\pm}$($\Sigma^0$) final 
states, the $\Sigma^*$ exchanges and the excited hyperon exchanges, 
of which the formulae have been given in Ref. 5.  The excited 
hyperons in this framework are treated as degenerate so that their total 
contributions can be written in a compact form in the quark model.  This
is a good  approximation since the contributions from the u-channels 
resonances are not sensitive to their precise mass positions.   
The transition amplitude ${\cal M}_s$ in Eq. 1 is
\begin{equation}\label{23}
{\cal M}_s=\sum_R \frac {2M_R}{s-M_R(M_R-i\Gamma_R ({\bf q}))}
e^{-\frac {{\bf k}^2+{\bf q}^2}{6\alpha^2}} {\cal O}_R,
\end{equation}
where the resonance $R$ has the mass $M_R$ and total width $\Gamma_R$,
 ${\bf k}$ and ${\bf q}$ are the momenta of  incoming photons
and outgoing mesons, and $\sqrt {s}$ is the total energy of the
system.  The operator ${\cal O}_R$ in Eq. \ref{22} depends on the 
structure of resonances, and it is divided into two parts:
 the s-channel resonances below 2 GeV and those above 2 GeV that
could be regarded as continuum contributions.
The electromagnetic transitions of the s-channel baryon resonances and
their meson decays have been investigated extensively in the quark 
model\cite{cko,LY,simon,close} in terms of the helicity and
the meson decay amplitudes.  These transition amplitudes for s-channel
resonances below 2 GeV
have been translated into the standard CGLN\cite{cgln} 
amplitudes in Refs. 5 and 6 for the proton target and
14 for the neutron target in the harmonic oscillator basis.
The advantage of the standard CGLN variables is that the kinematics
of the meson photoproductions has been thoroughly 
investigated\cite{tabakin},  the various observables of the meson 
photoproductions could be easily evaluated in terms of these amplitudes.
Those resonances above 2 GeV are treated as degenerate, since few 
experimental information is available on those resonances.
Qualitatively, we find that the resonances with higher partial waves
have the largest contributions as the energy increases.  Thus, we
write the total contributions from the resonances belonging to the same 
harmonic oscillator shell in a compact form, and the mass and total 
width of the high spin states, such as $G_{17}(2190)$ for $n=3$
harmonic oscillator shell, are used.

If we assume that the relative strength and phases of each term in s-,
u- and t-channels are determined by the quark model wavefunction
with exact $SU(6)\otimes O(3)$ symmetry, and the masses and decay widths
of the s-channel baryon resonances are obtained from the recent particle
data group\cite{pdg94},  there are four parameters in this calculation; 
the coupling constant $g_{KN\Sigma}$ or $g_{\eta NN}$, the constituent
 quark masses 
$m_q$ for up or down quarks and the strange quarks, and the parameter 
$\alpha^2$ from the harmonic oscillator wavefunctions in the quark
model.  The quark masses $m_q$ and the parameter $\alpha^2$ are
well determined in the quark model, they are
\begin{eqnarray}\label{22}
m_{u}=m_{d}=0.34 & \quad & \mbox{GeV} \nonumber \\
m_s=0.55 & \quad & \mbox{GeV}  \nonumber \\
\alpha^2=0.16 & \quad & \mbox{GeV}^2 .
\end{eqnarray} 
This leaves only {\bf one} free parameter, the coupling constant 
$\alpha_{K Y N}$ or $\alpha_{\eta NN}$, to be determined in the 
calculation.  The one parameter evaluation\cite{mawx} for all four 
isospin channels of the reaction $\gamma N\to K\Sigma$ shows an 
excellent overall agreement with the few available data in both 
differential and total cross sections.  It represents a dramatic 
improvement over the similar calculations in the isobar 
model\cite{mart}.  Our investigation in the $\eta$ 
photoproductions\cite{zpli952} also showed that the one parameter 
evaluation has produced very good agreement with the experimental
data well beyond the threshold region. 

An important feature in the chiral quark model approach is 
that it relates the photoproduction data directly to the spin flavor 
structure of the s-channel resonances. This is particularly the case for 
the S-wave resonances in the second resonance region, $S_{11}(1535)$
and $S_{11}(1650)$.  The recent data\cite{krusch} for the $\eta$ 
photoproduction in the threshold region from MAMI provides us
 more systematic information near the threshold region with much 
better energy and angular resolution.  Thus, it enables us to 
determine the properties of the $S_{11}(1535)$ resonance more precisely.
One property of the $S_{11}(1535)$ resonance, determined from the $\eta$
photoproduction, is given by the quantity $\xi$,
\begin{equation}\label{1}
\xi=\sqrt{ \frac {M_Nk \chi_{\eta N}}{qM_R\Gamma_T}}A_{\frac 12}
\end{equation}
where $M_N$ ($M_R$) denotes the mass of the nucleon (resonance),
$k$ and $q$ correspond to the momenta of the incoming photon and the
outgoing meson $\eta$, $\chi_{\eta N}$ is the branching ratio of the
resonance to the $\eta N$ channel, and $\Gamma_T$ and $A_{\frac 12}$
are the total width and the helicity amplitude for the resonance.
A study\cite{muko} by the RPI group shows that this quantity
obtained from the experimental data is model independent, and thus
should be calculated in theoretical investigations.
This quantity is given by an analytical form in the quark model,
\begin{eqnarray}\label{2}
\xi=\sqrt{\frac {\alpha_{\eta}\alpha_e\pi(E^f+M_N)}{M_R^3}}\frac
{C_{S_{11}(1535)}k}{6\Gamma_T}\left [\frac
{2\omega_{\eta}}{m_q}-\frac {2{
q}^2}{3\alpha^2}\left (\frac {\omega_{\eta}}{E^f+M_N}+1\right )\right
] \nonumber \\
\left (1+\frac { k}{2m_q}\right )e^{-\frac {{ q}^2+{
k}^2}{6\alpha^2}},
\end{eqnarray}
where $\omega_{\eta}$ and $E^f$ are the energies of the outgoing
$\eta$ meson and the nucleon.
The coupling of the $S_{11}(1535)$ to $\eta N$ in Eq. \ref{2} is
determined by the $\eta NN$ coupling constant $\alpha_{\eta}$.  This 
provides a consistency condition that must be checked in any microscopic
model of baryon
decay amplitudes,  otherwise, the overall agreement with data from
meson photoproduction would be lost. The coefficient $C_{S_{11}(1535)}$
is equal to unity in the naive $SU(6)\otimes O(3)$ quark model.
Thus the quantity $C_{S_{11}(1535)}-1$ measures a deviation of the resonance
wavefunction from the underlying  $SU(6)\otimes O(3)$ symmetry.
Both the $S_{11}(1535)$ and $S_{11}(1650)$ resonances
show a strong configuration mixing in more sophisticated models\cite{ik}.

By treating the coupling constant $\alpha_\eta$, the coefficient
$C_{S_{11}(1535)}$ and the total decay width $\Gamma_T$ as free parameters
and fitting them to the experimental data, we find\cite{zpli952}
\begin{eqnarray}\label{3}
\Gamma_T & = & 198 \quad \mbox{MeV} \nonumber \\
C_{S_{11}(1535)} & = & 1.608 \nonumber \\
\alpha_{\eta} & = & 0.435,
\end{eqnarray}
which gives an excellent fit to the recent MAMI data\cite{krusch}.
The above results give
\begin{equation}\label{31}
\xi = 0.220 \quad \mbox{GeV}^{-1}.
\end{equation}
This value is in good agreement with results of the RPI group\cite{muko},
 which used an effective Lagrangian approach  to fit both old data 
sets and new data from the Mainz group.  An extraction of the helicity
amplitude $A^p_{1/2}$ from the quantity $\xi$ depends on the $\eta N$
branching ratio $\chi_{\eta N}$,  which is not precisely known at present.
One could use, as a guide,
the result from a recent coupled channel analysis by
Batini\'c {\it et al}\cite{svarc}. There the
branching ratios to $\eta N$ and $\pi N$ channels
\begin{equation}\label{32}
\chi_{\eta N} =0.63 \;\; {\rm and} \;\;  \chi_{\pi N} =0.31,
\end{equation}
were found for the $S_{11}(1535)$ resonance, the latter being in
in good agreement with a result from the VPI group\cite{VPI}.
These lead to the helicity amplitude
\begin{equation}\label{4}
A^p_{1/2}=98.9 \quad 10^{-3} \; \mbox{GeV}^{-1/2}.
\end{equation}
However, the total width $\Gamma_T$ for the resonance $S_{11}(1535)$ varies significantly when extracted from recent partial
wave analyses\cite{VPI,svarc} and the $\eta$ photoproduction
data\cite{krusch}.  Thus, the helicity amplitude $A^p_{1/2}$ could not 
be determined reliably at present.

The constant $C_{S_{11}(1535)}=1.606$ in Eq. \ref{3} signals a large 
deviation mixing from the $SU(6)\otimes O(3)$ symmetry for the resonance
$S_{11}(1535)$, and it is unlikely that the configuration mixing effects
would account for such a large discrepancy.  This problem has been known 
for some time in the quark model.  The helicity amplitude $A^p_{1/2}$ from
quark model calculations has\cite{close,simon} remained near
150 $10^{-3}$ GeV$^{-\frac 12}$, while the branching ratio for the
$S_{11}(1535)$ resonance decaying to $\eta N$ is too
small\cite{zpli952,simon}.  It has been suggested\cite{chiral} that
a quasi-bound $K\Sigma$ state with properties remarkably similar to 
the resonance $S_{11}(1535)$ may be responsible for the large 
$\eta N$ branching ratio.  Such a scenario is unlikely since the
data\cite{stoler} for the electromagnetic transitions, in particular the $Q^2$ 
dependence of the helicity amplitude $A^p_{1/2}$, indicate that the 
resonance $S_{11}(1535)$ should be dominantly a $q^3$ state at higher 
$Q^2$ according to the pQCD counting rule\cite{carlson}.  In Ref. 27, 
we show that there is a considerable experimental evidence for a third
$S_{11}$ resonance with mass $1.712$ GeV and width 
$\Gamma_{S_{11}}=0.184$ GeV.  The recent multi-channel analysis by Dytman {\it
et al}\cite{dytman} also suggests the existence of this resonance, 
although the evidence is still weak.  This third $S_{11}$ resonance, 
if it exists in nature, can not be accommodated by the quark model, 
because its mass is near the other two known $S_{11}$ resonances, 
$S_{11}(1535)$ and $S_{11}(1650)$, and a quasi-bound $K\Sigma$ or 
$K\Lambda$ state strongly
mixed with the $q^3$ quark state would be a likely outcome.  The partial 
wave analysis by Deans\cite{dean} {\it et al}  suggests that the coupling
 of this resonance to $K\Sigma$ would be very strong.  Thus, the kaon 
production experiments, such as $\pi N\to KY$ and $\gamma N\to KY$, 
would be very important in establishing its existence.  Moreover, our 
investigation\cite{mawx} in $\gamma N\to K\Sigma$ shows that the contributions
from this resonance should be further enhanced by the threshold effects.
However, incorporating the contributions from this state requires 
more elaborate modeling and more precise data, which remains to
 be investigated.

Another advantage of the quark model approach for meson photoproductions
is that it can be extended to the photoproductions of heavy mesons, such
as the $\eta'$ photoproductions.  An interesting prediction\cite{etap}
 from the quark model emerges for the $\eta'$ photoproduction; 
the threshold behavior of the $\eta'$ photoproduction is dominated by the 
off-shell contributions from the s-wave resonances in the second resonance 
region,  which can be tested in the future CEBAF experiments\cite{rechi}.
In Fig. 1, we show the result of our calculation for the total 
cross sections of the $\etap$ photoproduction off the proton target.  
The coupling constant $\alpha_{\etap NN}$ is $0.35$ from the 
fit to a few total cross section data, which is indeed 
consistent with the recent studies\cite{etap1} suggesting it to 
be small.  Of course, there is a large uncertainty due to the 
poor quality of the data. Our result does not exhibit the 
dominance of any particular resonance around 2 GeV region suggested 
by the RPI group\cite{rpi1}.  This can be understood by the relative 
strength of the CGLN 
amplitudes ${\cal O}_R$ in Eq. \ref{2} between the s-wave resonances
in the second resonance region and the resonances around 2.0 GeV in 
the quark model.  There are two $S_{11}$ resonances with isospin 1/2
in the second resonance region.  The operator $O_R$ in Eq. \ref{23} 
for these two resonances are proportional to the quantity $\xi$ 
in Eq. \ref{2}, in which the leading term does not depend on the
outgoing meson momentum ${\bf q}$.  On the other hand, the 
$S$ or $D$ wave resonances around 2 GeV belong to $n=3$ in the harmonic 
oscillator basis.  The operator ${\cal O}_R$ for the $n=3$ resonances is
\begin{eqnarray}\label{4}
{\cal O}_{n=3} = -\frac {1}{12m_q}i\vsig \cdot {\bf A}\vsig 
\cdot (\vep\times {\bf k})\left (\frac {{\bf k}
\cdot {\bf q}}{3\alpha^2}\right )^3 \nonumber \\
+\frac 1{6}\left [\frac {\omega_{\etap} k}{m_q}\left (1+\frac 
{k}{2m_q}\right )\vsig \cdot  \vep  + \frac
{k}{\alpha^2}\vsig\cdot {\bf A}\vep\cdot
{\bf q}\right ]\left (\frac {{\bf k}\cdot {\bf q}}
{3\alpha^2}\right )^2 \nonumber \\ +\frac {\omega_{\etap} k}{9\alpha^2 m_q}
\vsig\cdot {\bf k}\vep\cdot {\bf q} \left (\frac {{\bf k}\cdot {\bf q}}{3
\alpha^2} \right ).
\end{eqnarray}
Because the spatial 
wavefunctions for the $S$ and $D$ wave resonances are orthogonal to that
of the $S_{11}(1535)$,  the dependence on ${\bf k}$ and ${\bf q}$ of the 
CLGN amplitudes for the $S$ and $D$ wave resonances around 2 GeV should 
be very different from those for $S_{11}(1535)$ and $D_{13}(1520)$.  
Indeed, Eq. \ref{4} shows that the amplitude for the $S$ and $D$ wave 
resonances with $n=3$ is at least proportional to ${\bf q}^2$ comparing
to the ${\bf q}$ dependence of the amplitude of the $S_{11}(1535)$
in Eq. 5.  Thus, the $S$ and $D$ wave resonances around 2 GeV give 
little contribution to the $\etap$ productions in the threshold region.
Moreover, Eq. \ref{4} represents the 
sum of every resonance with $n=3$, and the $G$ wave resonance has the 
largest amplitude among these resonances.  The magnitude of the CGLN
amplitudes for $S$ and $D$ wave resonances is even smaller that that
for the $S_{11}(1535)$, since the magnitude of Eq. \ref{4} is about 
10 times smaller than that of Eq. 5.  The results in Fig. 1
represents an important prediction of the quark model, since the relative
strength and phases of the CGLN amplitudes for each s-channel resonance
are determined by the quark model wavefunctions.  It also shows the importance
of the $S_{11}$ resonances in the second resonance region in the meson
photoproductions, in particular the $K$, $\eta$ and $\etap$ productions.

In summary, the quark model approach represents a 
significant advance in the theory of the meson photoproductions.
It introduces the quark and gluon degrees of freedom explicitly, 
which is an important step towards establishing the connection between 
the QCD and the reaction mechanism.  It highlights the dynamic roles 
by the s-channel resonances,  in particular the roles of the $S_{11}$
resonances in the threshold region of the $K$, $\eta$ and $\etap$
 photoproductions.  Perhaps more importantly, this approach provides an 
unified description of the meson photoproductions.  The challenge
for this approach would be to go one step further so that the 
quantitative descriptions of meson photoproductions, in particular 
the polarization observables that are sensitive to the detail structure
of the s-channel resonances, could be provided.  This investigation 
is currently in progress.

The author wishes to thank the organizers of this workshop for inviting
me to give this talk.  The financial support from Peking University, and
the U.S. National Science Foundation grant PHY-9023586 are also gratefully
acknowledged.

\begin{figure}
\centering
\epsfbox{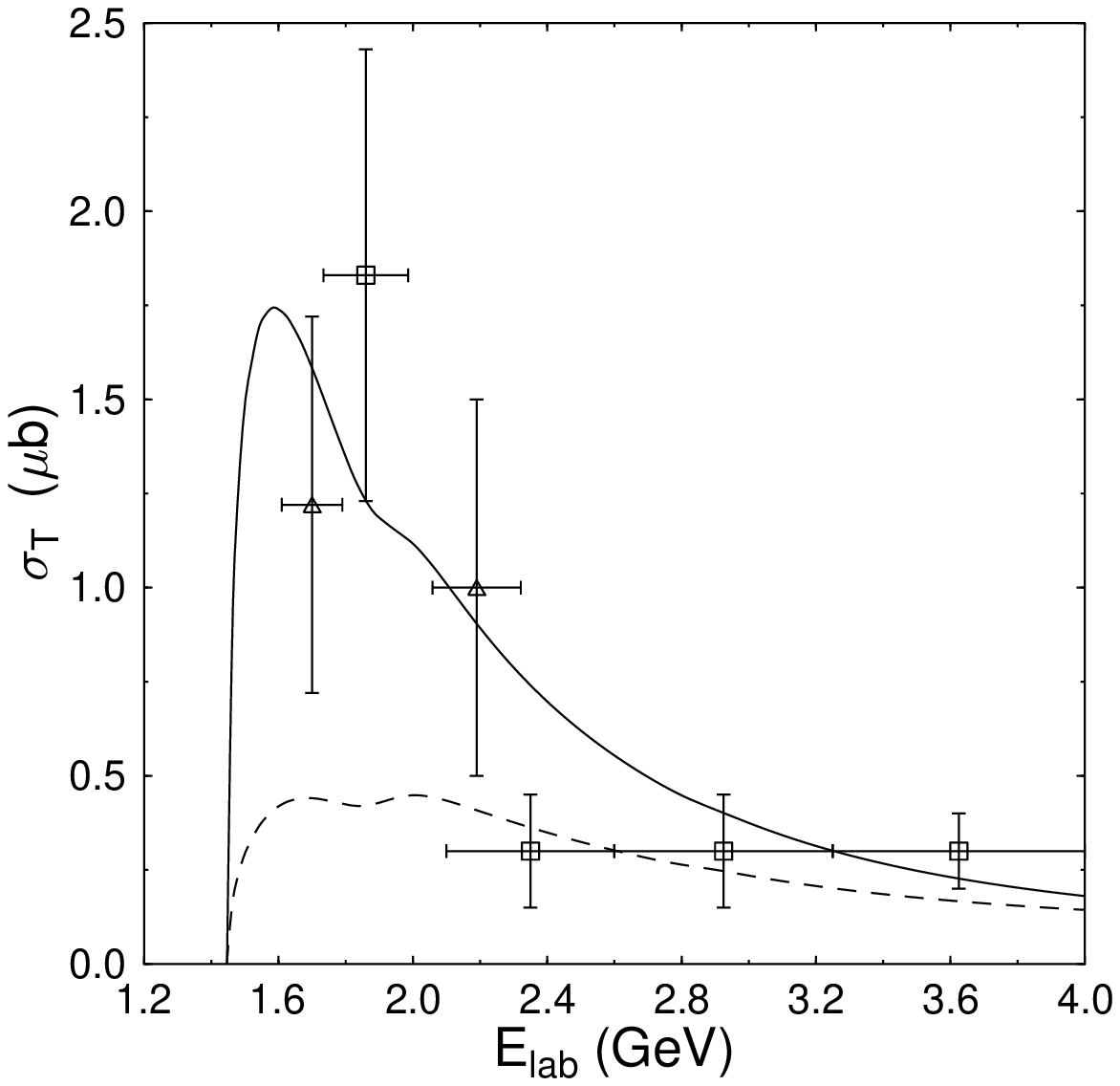}
\caption[fig1]{ The total cross sections for $\gamma + p\to \eta' + p$.
The difference between the solid and dashed lines represents
the importance of the contribution from the resonance $S_{11}(1535)$.
 The experimental data are from Refs. 32 (triangle) 
and 33 (square).}
\label{fig1}
\end{figure}

\end{document}